\documentclass[conference]{IEEEtran}
\IEEEoverridecommandlockouts

\makeatletter
\def\ps@headings{%
\def\@oddhead{\mbox{}\scriptsize\rightmark \hfil \thepage}%
\def\@evenhead{\scriptsize\thepage \hfil \leftmark\mbox{}}%
\def\@oddfoot{}%
\def\@evenfoot{}}
\makeatother
\usepackage{graphicx, subfigure, epsfig}
\usepackage{url}
\usepackage{algorithm}
\usepackage{algorithmic}
\usepackage{amsmath}
\usepackage{amssymb,amsfonts}

\newcommand {\beq} {\begin{equation}}
\newcommand {\eeq} {\end{equation}}

\begin{document}

\title{Diffusive Logistic Model Towards Predicting Information Diffusion in Online Social Networks}

\author{\IEEEauthorblockN{Feng Wang, Haiyan Wang, Kuai Xu}
\IEEEauthorblockA{Arizona State University\\
Email: \{fwang25, haiyan.wang, kuai.xu\}@asu.edu}
}

\maketitle

\thispagestyle{empty}
\begin{abstract}
Online social networks have recently become an effective and innovative channel for spreading information and influence among hundreds of millions of end users. Many prior work have carried out empirical studies and proposed diffusion models to understand the information diffusion process in online social networks. However, most of these studies focus on the information diffusion in temporal dimension, that is, how the information propagates over time. Little attempt has been given on understanding information diffusion over both temporal and spatial dimensions. In this paper, we propose a Partial Differential Equation (PDE), specifically, a Diffusive Logistic (DL) equation to model the temporal and spatial characteristics of information diffusion in online social networks. To be more specific, we develop a PDE-based theoretical framework to measure and predict the density of influenced users at a given distance from the original information source after a time period. The density of influenced users over time and distance provides valuable insight on the actual information diffusion process. We present the temporal and spatial patterns in a real dataset collected from Digg social news site, and validate the proposed DL equation in terms of predicting the information diffusion process. Our experiment results show that the DL model is indeed able to characterize and predict the process of information propagation in online social networks. For example, for the most popular news with 24,099 votes in Digg, the average prediction accuracy of DL model over all distances during the first 6 hours is $92.08\%$. To the best of our knowledge, this paper is the first attempt to use PDE-based model to study the information diffusion process in both temporal and spatial dimensions in online social networks.

\end{abstract}

\section{Introduction}

Recent years have witnessed the explosive growth of online social networks that connect people in the digital world. Social networking sites such as Facebook, Twitter, and Digg provide innovative services through a wide variety of applications for hundreds of millions of users. One of the most important functionalities of online social networks is to effectively spread information such as latest news headlines, movie recommendations, etc, across the networks. This process is called information diffusion. 

Given the significant role online social networks have played in elections and crisis~\cite{Hughes09}, it has become increasingly urgent to gain a deep understanding of information diffusion process through online social networks. Furthermore, understanding the speed and coverage of an information spreading over online social networks also provides interesting insights on user interactions. However, understanding the process of information diffusion over online social networks is a daunting task due to the growing scale, dynamics, and complexity of these networks.

Most prior work on online social networks have focused on the understanding of network structure, user interactions, and traffic characteristics~\cite{JiangIMC10, Schneider:IMC09, Benevenuto:IMC09, Nazir:IMC09, Nazir:IMC08}. Several work have studied the characteristics of information diffusion over online social networks using empirical approaches~\cite{LermanICWSM10, SteegPS11, YeCSI10, ChaWOSN08, Cha09} to measure the number of votes or comments for an information over time, or the average number of social links the information has traveled, etc. A few recent effort use mathematical models to predict information diffusion over a time period in online social networks~\cite{YangICDM10,GhoshWSDM11}. However, little attempt has been given on understanding and modeling information diffusion in both temporal and spatial dimensions.

This paper proposes a novel Diffusive Logistic (DL) model to study both temporal and spatial patterns of information diffusion process in online social networks. Specifically we are interested in answering the {\em spatio-temporal diffusion problem}: for a given information $m$ initiated from a particular user called {\em source $s$}, after a time period $t$, what is the density of influenced users at distance $x$ from the source? An influenced user is an user who has actively voted or liked the information. Solving this spatio-temporal diffusion problem could not only shed light on the detailed process of information diffusion, but also help predict the spreading patterns of similar information in the future.

In our DL model, we innovatively translate the information diffusion process in online social networks into two separate processes: growth process and diffusion process. Growth process represents information spreading among users with the same distance from the source, and is measured with the simple and well-known logistic model~\cite{Murray1989}. Diffusion process is the process through which information randomly spreads among users at different distances from the source. We measure diffusion process with Fick's law of diffusion~\cite{Murray1989}. We also introduce two metrics, {\em friendship hops} and {\em shared interests} to measure distance in online social networks. Furthermore, we theoretically analyze our model and prove that the DL model has {\em unique property} and {\em strictly increasing property}. These two properties make the DL model a good candidate for modeling the density of influenced user in online social network context.

We carry out empirical studies to understand the temporal and spatial patterns of information diffusion in a real dataset collected from Digg, a major social news aggregation site. The dataset consists of millions of votes on top news stories on Digg site during June 2009, and the friendship links among thousands of users who have voted these stories. We find out that both friendship hops and shared interests are good indicators of distance in Digg networks. In both cases, the density of influenced users shows consistent patterns of evolving over time. This affirms our choice of a PDE-based DL equation to model information diffusion in online social networks. Furthermore, we validate the proposed DL model on the Digg dataset. The experiment results show that by constructing the proper initial condition and parameters, the DL model can effectively predict the density of influenced users for a given distance and a given time for both distance metrics. For example, for the most popular news with 24,099 votes in Digg, the average prediction accuracy of DL model over all distances (with friendship hop as distance metric) during the first six hours is $92.08\%$. This further verifies the practicality of the DL model.

To the best of our knowledge, this paper is the first attempt to study the spatio-temporal diffusion problem in online social networks and propose PDE-based models for characterizing and predicting the temporal and spatial patterns of information diffusion over online social networks. It is worth noting that our proposed DL model is a PDE model which has been widely used in mathematical biology, sociology, economics, and physics to model multi-dimensional systems. By introducing PDE to online social networks and validating its effectiveness with real datasets, we demonstrate the clear similarity among spatial population dynamics in biology, heat conduction through metal bars in physics, and the information diffusion in online social networks. The contributions of this paper include:
\begin{itemize}
\item We introduce the 2D diffusion problem to understand information diffusion in online social networks;
\item We innovatively abstract the diffusion process and introduce DL equation to model diffusion process in online social networks;
\item We present the temporal and spatial patterns of information diffusion in a real dataset collected from a major online social news aggregation site;
\item We validate the DL model and evaluate its performance by matching its prediction with the real dataset.
\end{itemize}

The remainder of this paper is organized as follows. Section II introduces the DL model for modeling information diffusion in online social networks, gives theoretical analysis of the model and describes the construction of the initial density function. Section III presents empirical studies of the temporal and spatial patterns in the Digg datasets, and validates the effectiveness of the DL model in terms of prediction accuracy. Section IV gives a brief literature review on related work, and Section V concludes the paper and outlines our future work.

\section{Diffusive Logistic Model}

In this section, we first define two metrics to quantify the distance between two users in online social networks, then describe a PDE-based Diffusive Logistic (DL) model to characterize information diffusion process. Subsequently we analyze the properties of DL model and give guidelines on constructing initial density function and selecting parameters of DL model.

\subsection{Distance Metrics}
In this paper, we study the information diffusion in two dimensions: temporal and spatial dimensions. Before we present the mathematical model, we need to give a clear definition
of distance which is correspondent to the spatial dimension. We propose to quantify the distance from two perspectives: {\em friendship hops} and {\em shared interests}.
A natural metric of distance between two users is the length of the shortest path, measured by the number of hops from one user to another in the social network graph. We call this distance metric {\em friendship hops}.

In online social networks, {\em interest} is another metric. For two users without any direct friendship, if they have shown similar interests, it is highly possible that information will diffuse between them since
an online social network provides a convenient channel for sharing and communicating beyond friendship. Therefore, we introduce an alternative metric called {\em shared interests} to measure the distance between two users. Given two social networks users $a$ and $b$, let $C_{a}$ and $C_{b}$ denote the set of contents both
users $a$ and $b$ have interacted with (In the context of Digg networks, $C_{a}$ represents all the news stories the user $a$ have voted or digged), the {\em shared interests} $d_{a,b}$ between users $a$ and $b$ is
defined as:
\begin{equation}
d_{a,b} = 1 - \frac{C_a \cap C_b}{C_a \cup C_b}
\end{equation}
where $C_a \cup C_b$ is the number of the total contents that
either user $a$ or user $b$ has interacted with, and $C_a \cap C_b$
is the number of the shared contents that both users $a$ and $b$
have interacted with. Essentially the interest distance
quantifies the degree of the shared interests among two
users. An information originating from the source $s$ is more likely
to influence users who have small shared interests distances than users with large ones.

\subsection{Diffusive Logistic Model}
Information and influence spread dynamically in many different ways through the complex online social networks. It is challenging to tell the topology of the underlying diffusion networks~\cite{GomezRodriguez:2010}. In general, the information can spread along social links between users that are direct friends and can also spread in a random way among users without direct social links.  For example, in Digg networks, information spreading happens when a follower votes for news submitted by his followee. In addition, a user, who is not a follower of the users who have voted a news, can also votes for the same news after the news is promoted to the front page, or listed by the search functions of the Digg site. Hence, information propagation also happens between these two voters who are not direct friend. In this section, we propose a PDE-based Diffusive Logistic model to characterize information diffusion in dynamic and complex online social networks. We propose an innovative approach to abstractly divide the diffusion into two processes which can be respectively modeled with mathematical models widely used in mathematical biology, sociology, economics, and physics domains.

\subsubsection{Model Heuristic}
Let $U$ denote the user population in an online social network, and $s$ is the source of an information. For social network users, based on their distances to the source,
the user population $U$ is breakdown into a set of
groups, i.e., $U = \{U_1, U_2, ... U_i, ..., U_m\}$, where $m$ is the maximum distance. The group $U_x$ consists of users that have distance $x$ to the source.

As information propagates through social networks, some users express their interest in the information by commenting, voting, forwarding, digging or other
activities. We call such users as {\em influenced users} of the information.
Let $I(x,t)$ denote the density of influenced users at distance $x$ during time $t$, that is, $I(x,t)$ reflects the ratio of the number of influenced users with a distance of
$x$ at time $t$ over the total number of users in $U_x$. The value of $I(x,t)$ depends on two diffusion processes. First, the users in $U_{y}$ where $y \neq x$ can influence
those in $U_x$ through direct or indirect friendship links that can be either uni-directional or bi-directional. We call this process {\em diffusion process}. We assume that this diffusion is in the manner of random walk, that is, users at distant $x$ randomly influence the users at a different distance. Secondly, the users in $U_x$ could also influence each other. In online social networks, social triangles, also called triads formed by high clustering of users, are very common in online social networks. Therefore, it is possible that two users of the same distance from the source are also friends themselves. We call this process {\em growth process}.

\begin{figure}[htbp]
    \centering{
    \includegraphics[width=3.0in]{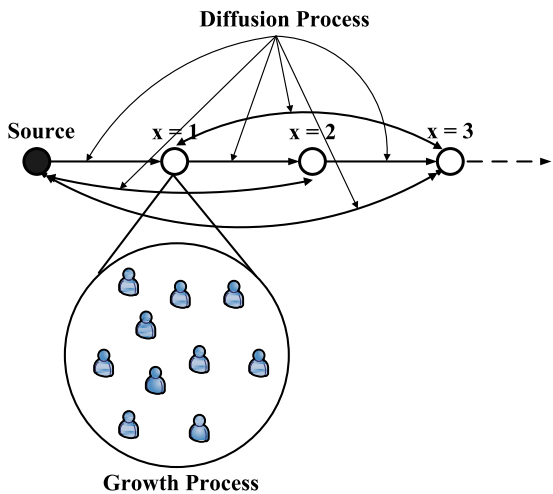}}
 \caption{Information spread processes in online social networks}
\label{fprocess}
\end{figure}

Figure~\ref{fprocess} illustrates the diffusion process and growth process. The circle $x=i$ represents all the users that have distance $i$ from the source. It contains many users that spread information among themselves. Users at a certain distance can spread information to users at other distances in a random fashion.

\subsubsection{Model Description}
The growth process can be modeled with simple logistic model~\cite{Murray1989}. Logistic model is widely used to
model the population dynamics where the rate of reproduction
is proportional to both the existing population and the amount
of available resources. It has also been used to describe various
population dynamics and predict growth of bacteria and tumors
over time, etc~\cite{Murray1989}. Logistic equation is defined as follows.
Denoting with $N$ the population at time $t$, $r$ the intrinsic
growth rate and $K$ the carrying capacity which gives the upper
bound of $N$, the population dynamics are governed by:

\begin{equation}
N^{\prime} = rN(1-\frac{N}{K})
\end{equation}
where $N^{\prime}$ is the first derivative of $N$ with respect to $t$.

In the context of online social networks, we are interested in modeling the impact of the user influence
within the same group $U_x$ on the growth of $I(x,t)$, the density of influenced users at the
distance $x$ during time $t$. Hence, the growth process is modeled
as:
\begin{equation}
\frac{\partial I(x,t)}{\partial t} = rI(x,t)(1-\frac{I(x,t)}{K})
\end{equation}

The diffusion process can be naturally modeled with Fick's law of diffusion~\cite{Murray1989}, which has been used
to measure the diffusion of heat in a metal as well as the diffusion of chemical specifies.
Based on Fick's law, the random spreading of information among user groups at different distances can be measured with $d \frac{\partial^2 I}{\partial x^2}$
if the diffusion rate is set as $d$. \\

Combining the growth process and diffusion process together,
we derive the diffusive logistic equation as follows:

\begin{equation}\label{eq1}
\begin{split}
&\frac{\partial I}{\partial t}=d \frac{\partial^2 I}{\partial x^2}+r I(1-\frac{I}{K})\\
& I(x, 1) = \phi(x), \;\;l\leq x \leq L  \\
&\frac{ \partial I}{\partial x}(l,t)=\frac{ \partial I}{\partial x}(L,t)=0, \;\;  t \geq 1 \\
\end{split}
\end{equation}

where
\begin{itemize}
\item $I$ (interchangeable with $I(x,t)$) represents the density of influenced users with a distance of $x$ at time $t$;
\item $d$ represents the diffusion rate measuring how fast the information travels across distances in social networks;
\item $r$ represents the intrinsic growth rate of influenced users with the same distance, and measures how fast the information spreads within the users with the same distance;
\item $K$ represents the carrying capacity, which is the maximum possible density of influenced users at a given distance;
\item $L$ and $l$ represent the lower and upper bounds of the distances between the source $s$ and other social network users;
\item $\phi(x)$ is the initial density function. It is non-negative and not identical to 0. Each information has unique $\phi$ which can be constructed from the initial phase of spreading;
\item $\frac{ \partial I}{\partial t}$ represents the first derivative of $I$ with respect to time $t$;
\item $\frac{ \partial^2 I}{\partial x^2}$ represents the second derivative of $I$ with respect to distance $x$.
\end{itemize}

$\frac{ \partial I}{\partial x}(l,t)=\frac{ \partial I}{\partial x}(L,t)=0$ is the Neumann boundary condition~\cite{Murray1989},
which means no flux of information across the boundaries at $x=l,L$. This is true for online social networks since information spreads within the networks.

\subsection{Theoretical Analysis of DL Model}
\vspace{5pt}

\noindent {\em Definition 1}: Lower time-independent solution $u(x)$ of Equation (\ref{eq1}) satisfies
\begin{equation}\label{eq12}
\begin{split}
&d u''+r u(1-\frac{u}{K}) \geq 0\\
& u(x) \leq \phi(x),  \;\;l\leq x \leq L  \\
& u'(l)=u'(L) = 0, \;\;
\end{split}
\end{equation}
where $u$ is a function of distance $x$, $u''$ represents the second derivative of $u(x)$ and $u'$ represents the first derivative of $u(x)$.  \\
A \emph{upper time-independent solution }$u(x)$ of Equation~(\ref{eq1}) can also be defined by reversing the inequality in Equation (\ref{eq12})

\vspace{5pt}
\noindent {\em Unique Property}: The DL model has a unique positive solution $I(x,t)$ and $0 \leq I(x,t) \leq K$. \\
\\
{\em Proof}: This is because Equation (\ref{eq1}) has two equilibria $I = 0$ and $I = K$ which are lower and upper solutions of Equation (\ref{eq1}) respectively. Based on Theorem 2.5.2 in~\cite{Pao1992}, the DL model has a unique solution between $0$ and $K$.
\hfill $\Box$

\vspace{5pt}

\noindent {\em Strictly Increasing Property}: The solution $I$ of Equation (\ref{eq1}) is strictly increasing with respect to time $t$ if initial density function $\phi(x)$ is a lower time-independent solution. \\
\\
{\em Proof}:  This property is a special case of a general results in  partial differential equations (see Lemma 5.4.1 in~\cite{Pao1992}).  In fact, from the comparison principle ~\cite{Pao1992}, $I$ always lies between $\phi(x)$ and $K$. Let $\delta>0$ and consider the difference of $I(x,t+\delta)-I(x,t)$. The comparison principle implies that $I(x,t+\delta)-I(x,t)\geq 0.$ In addition, if $\phi(x)$ itself is not a solution of  Equation (\ref{eq1}), then  $I(x,t)$ is strictly increasing with respect to $t$.
\hfill $\Box$

\vspace{5pt}

The unique property and strictly increasing property of the proposed DL model support
the application of this model to information diffusion process over online social networks,
because as the time goes, the density of influenced users always increases.
More importantly, the DL model could lead to a unique and strictly increasing solution for predicting the diffusion process.
The experiment results in the next section also verify these two important properties of the DL model.

\subsection{Initial Density Function Construction}

Now we present the method to construct the initial density function $\phi$ and provide some guidelines for model parameter selection.
In general, the initial function is constructed using the data collected from the initial stage
of information diffusion. Specifically, the function $\phi$ which is a function of distance $x$
captures the density of influenced user at distance $x$ at time $t = 1$.

The DL model has three requirements on the initial
function $\phi$: i) the function has to be twice continuous differentiable;
ii) the slopes at the left and right ends are zero, that is, $\phi'(l)=\phi'(L) = 0$;
and iii)
\begin{equation}\label{eq_init}
d \phi''+r \phi(1-\frac{\phi}{K}) \geq 0.
\end{equation}

In online social networks, it is only possible to observe discrete values for the
initial density function, because the distance $x$ is discrete.
To satisfy the first requirement of the model, we apply a simple and effective
mechanism available in Matlab cubic spline package, called {\em cubic splines
interpolation}~\cite{Gerald1994}, to interpolate the initial discrete data in constructing $\phi(x)$.
Using this process, a series of unique cubic polynomials
are fitted between each of the data points, with the stipulation that the obtained curve
is continuous and smooth. Hence $\phi(x)$ constructed by the cubic splines
interpolation is a piecewise-defined function and twice continuous differentiable.
After cubic splines interpolation, we simply set the two ends to be flat to satisfy
the second requirement, since in this way the slopes of the density function $\phi(x)$
at the left and right ends are zero. For the last requirement, we have
$\phi(x) \leq K$, since $K$ is the carrying capacity. Thus $r \phi(1-\frac{\phi}{K}) \geq 0$.
If $\phi$ is convex, then $d \phi'' \geq 0$ and Equation~\ref{eq_init}
holds\footnote{In our experiments with Digg data, most region of $\phi$ is convex using friendship hops as distance.}.
If $\phi$ is concave in some range, Equation~\ref{eq_init} will still hold, as long as
$K$ is relatively large and the diffusion rate $d$ is sufficiently smaller than growth rate $r$.

Parameters $r$, $d$, $K$ can be constants or functions of time $t$ and distance $x$.
In general, growth rate $r$ controls the gap between $I(x,t)$ and $I(x,t+1)$ and is usually a function of $t$. Diffusion rate $d$ controls the slope of $I$, and
carrying capacity $K$ controls the upper bound of $I$.

\section{Experiment Evaluations}
In this section, we present our findings of information diffusion characteristics in an anonymized Digg dataset~\cite{Digg2009} and evaluate the performance of the proposed DL model with the dataset. We'd like to answer two specific questions: first, what is the density of influenced users at distance $x$ from the source at time $t$; second, how to forecast and predict the density given density data collected at the initial stage of a news. We first describe the anonymized Digg dataset, then present the temporal and spatial characteristics of the dataset, and finally validate the DL model by comparing the predicted density with the actual observations in the dataset. \\

\subsection{Digg Data Set}
The dataset used in this study is collected from Digg, one of the most popular news aggregation sites. Users can submit links of news stories that they find in professional news sites and blogs to Digg, and can vote and comment on the submitted news. Digg users form friendship links through "following" each other. The first voter who brings the news to the Digg site, is called the initiator or source. There are two ways of information propagation in Digg: 1) A user can see the news submitted by the friends he follows and vote the news. After a user votes for a news, all his followers are able to see and vote on the news, and so on. 2) Once the news is promoted to the front page due to high popularity, the users, who do not friend with the initiator directly or indirectly, will also be able to view and vote for the news. A user can also discover news through search engines provided by the site and vote for it. The second approach of information propagation confirms our assumption of random walk of information spreading. Thus the Digg data provides an opportunity for us to study the impact of the friendship relationships and random walk of diffusion on the process of information spreading.

The datasets consist of $3553$ news stories that are {\em voted} (also called {\em digged}) and promoted
to the front page of {\tt www.digg.com} due to vote popularity during June 2009. In total, there are more than 3 millions votes cast on these news stories from over
$139,409$ Digg users. In addition, the datasets also include the directed {\em friendship} links among the Digg users who have voted these news stories. Based on these friendship links,
we construct a directed social network graph among these Digg users. For each of the news stories, the datasets include the user id
of all the voters during the collection period, and the timestamps when votes are cast. The time
granularity is in seconds. The timestamp and the social network graph provide critical data to study the temporal and spatial patterns of information propagation.

\subsection{Characterizing Temporal and Spatial Patterns of Information Diffusion}
To gain a better understanding of information diffusion in online social networks, we
characterize the temporal and spatial patterns of information diffusion process in Digg dataset.
Specifically we study how the information propagates over {\em time} in online social networks,
and examine the impact of the {\em distance} measured by {\em friendship hops} and {\em shared interests}. Since each news is independent, we demonstrate
the results of four representative news of different vote scales. Story $s1$ is the most popular news with 24,099 votes, story $s2$ is
the second most popular news with 8521 votes, story $s3$ is a news with 5988 votes, and story $s4$ is a news with 1618 votes.

\subsubsection{Distance Based on Friendship Hops}

\begin{figure}[htbp]
    \centering{
    \includegraphics[width=2.5in]{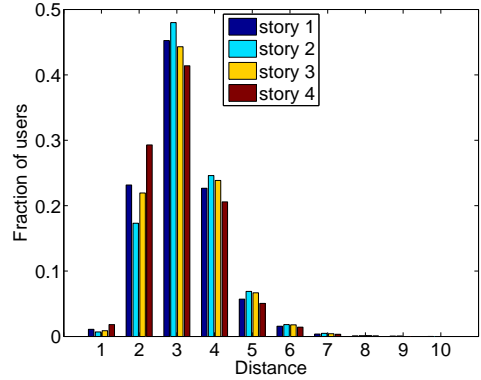}}
 \caption{Distribution of neighbors of the initiators of the four representative stories}
\label{fneighbor}
\end{figure}

\begin{figure*}[htbp]
  \centering
  \subfigure[Density of influenced users of s1]{
    \includegraphics[width=2.5in]{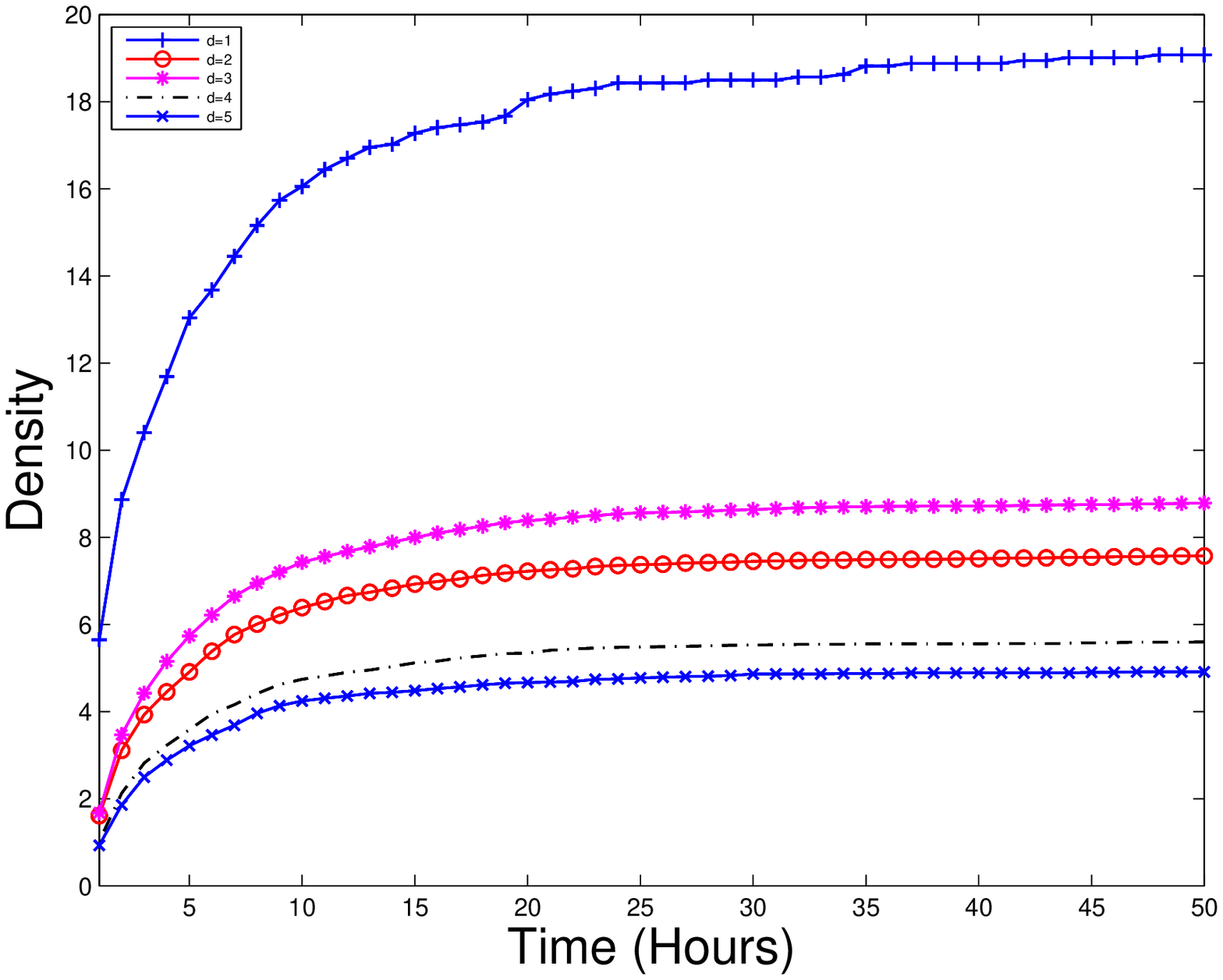}}
  \subfigure[Density of influenced users of s2]{
    \includegraphics[width=2.5in]{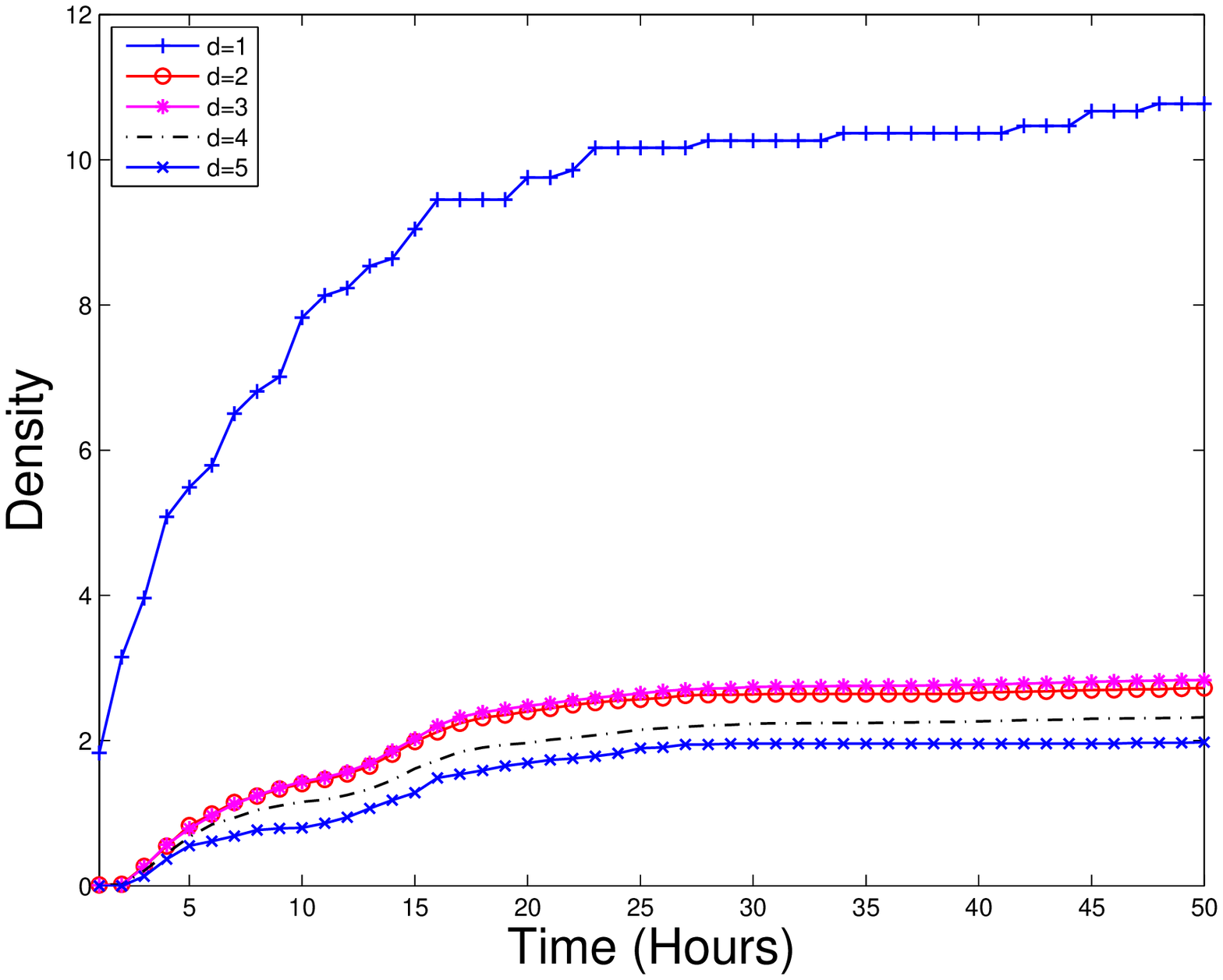}}
  \subfigure[Density of influenced users of s3]{
    \includegraphics[width=2.5in]{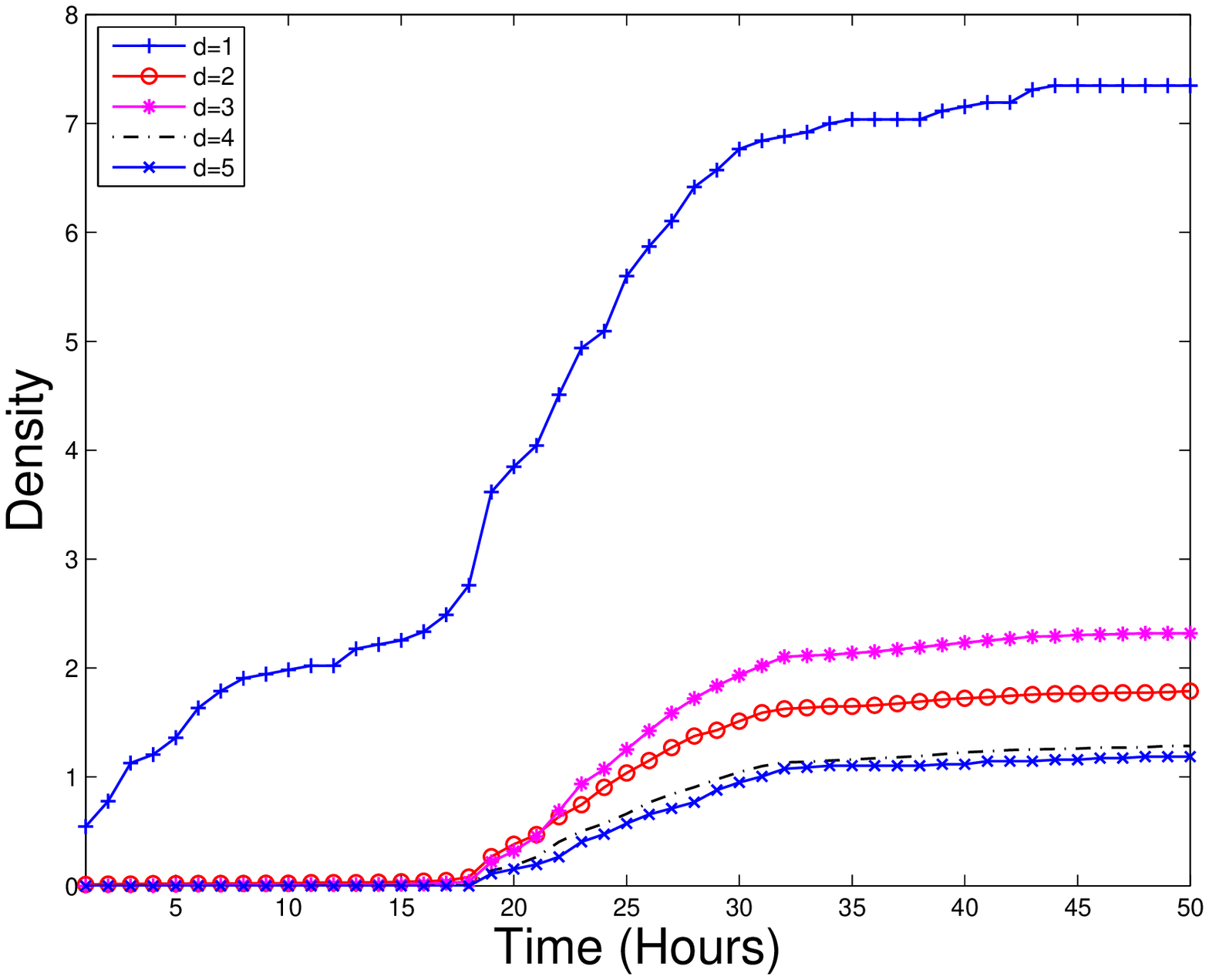}}
  \subfigure[Density of influenced users of s4]{
    \includegraphics[width=2.5in]{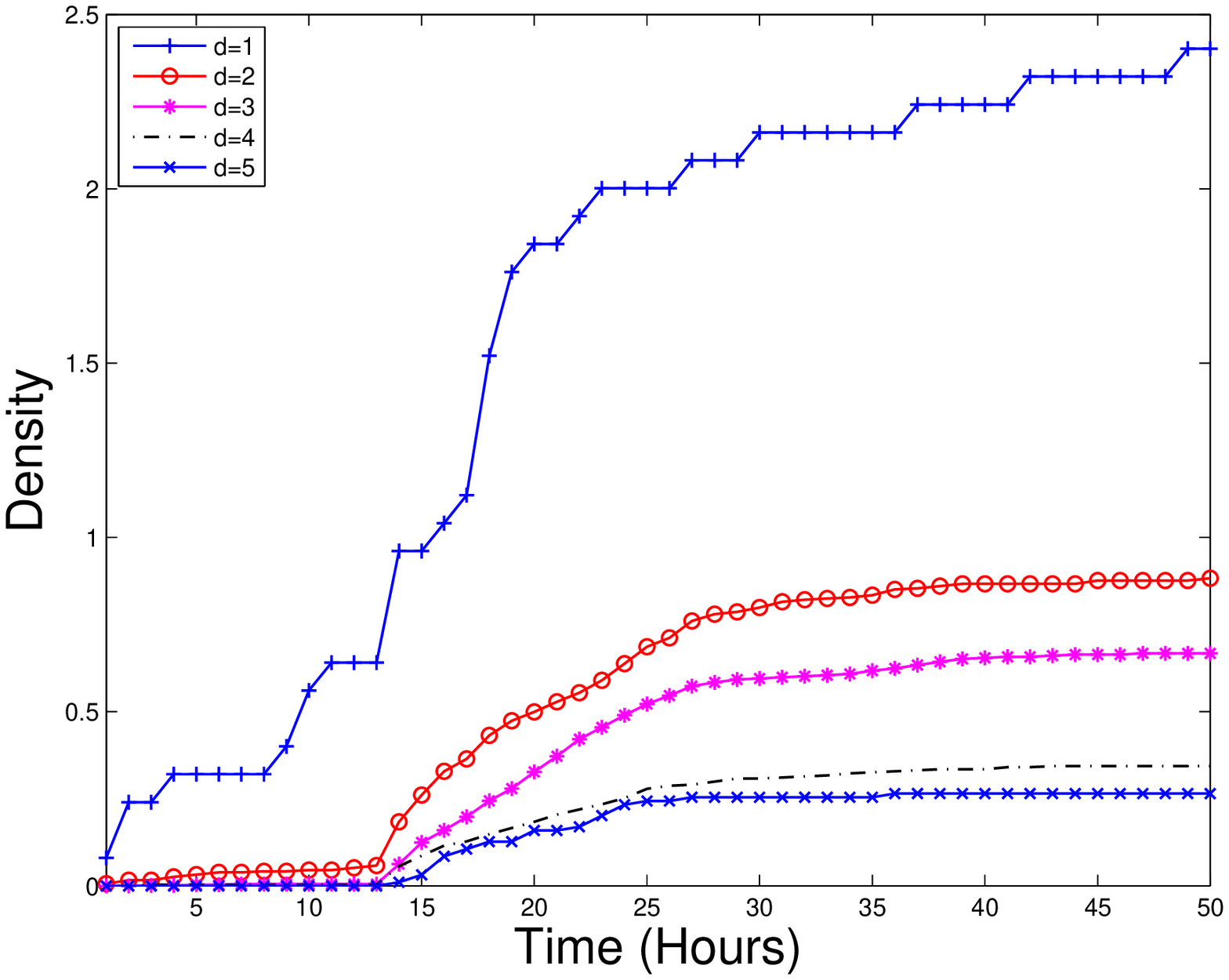}}
 \caption{Density of influenced users over 50 hours with friendship hops as distance}\label{fdensity}
\end{figure*}

As discussed in the previous section, we define the distance between the initiator and any
other user as the length (the number of friendship hops) of the shorted path from the
initiator to this user in the social network graph. Clearly, the direct followers of
the initiator have a distance of $1$, while their own direct followers have a distance
of $2$ from the initiator, and so on. Figure~\ref{fneighbor} shows the distance distributions
of the direct and indirect followers of the initiators of the four representative stories. An interesting observation is that the majority of
social network users have a distance of $2$ to $5$ from the initiators. For example, for all four stories,
the users of distance 3 account for more than $40\%$ of all the users reachable from the initiators. As the
distance increases from $6$ to $10$, the number of social networks users at these distances drops sharply. Due to the small size of users of distance $6$ to $10$, we only present the results for users of distance 1 to 5 in the remaining of this section.

Figure~\ref{fdensity}[a-d] illustrate the density of influenced users at 1 to 5 friendship hops from the initiators
over 50 hours for four stories. In the context of Digg social networks, we consider the users who have voted the news story as
influenced users. x-axis is the time, y-axis is the density of influenced users. Each line represents the density at a certain distance. The lowest line is the density at distance $1$.

There are five interesting observations:
\begin{itemize}
\item The density of influenced users evolves with time. Such a phenomena can be described by spatial-temporal PDE, which leads us to come up with the DL model;
\item Figure~\ref{fdensity}[a] shows that for story $s1$, the density of influenced users that are 3 hops away from the initiator is higher than that of users 2 hops away. This verifies that social links are not the only channel of information diffusion in Digg networks. Otherwise, if information only spreads along social links, then intuitively, as the distance increases, the density should decrease;
\item The density of influenced users at distance $1$ (represented by the top line) is significantly higher than that of users with hops greater than $1$. For story $s4$, the density of influenced users decreases as hops increases. This indicates that even there are other channels, the social links do play an important role in information diffusion in Digg networks;
\item Popular stories spread faster than less popular ones. For example, the density of influenced users of story $s1$ remains stable after about $10$ hours while story $s2$ (shown in Figure~\ref{fdensity}[b]) is stable after about $20$ hours;
\item A consistent observation over
all these sample stories is that after 50 hours, the densities of influenced users
for all distance remain stable, which suggests that the news stories are no longer "new" anymore.
\end{itemize}

\begin{figure}[ht]
  \centering{
    \includegraphics[width=3.0in]{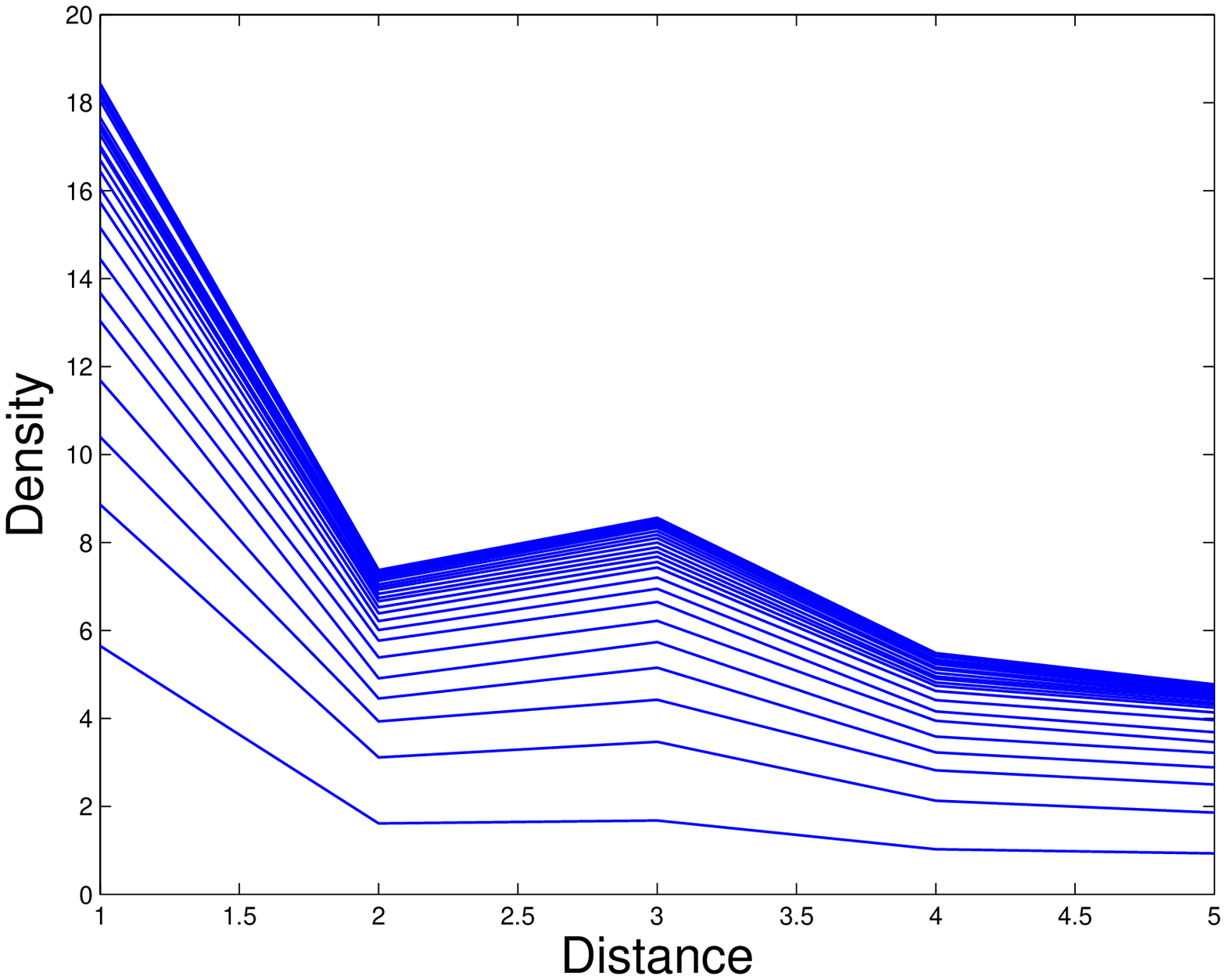}
  }
 \caption{Density of influenced users over 50 hours with friendship hops as distance} \label{f332346.hop.distance}
\end{figure}

Figure~\ref{f332346.hop.distance} illustrates the density of influenced users of the most popular story s1 from a different perspective. The x-axis is the distance, y-axis is the density of the influenced users, and each of the 50 lines represents the density at time $t$ where $t$ varies from 1 hour to 50 hours. It shows that as time passes by, the density of influenced users increases. However, the increment of densities at $t$ and $t+1$ decreases at time elapses. This leads us to choose a decreasing function of time $t$ for growth rate $r$ in the DL model.

\subsubsection{Distance Based on Shared Interests}
\begin{figure*}[htbp]
  \centering
  \subfigure[Density of influenced users of s1 ]{
    \includegraphics[width=2.5in]{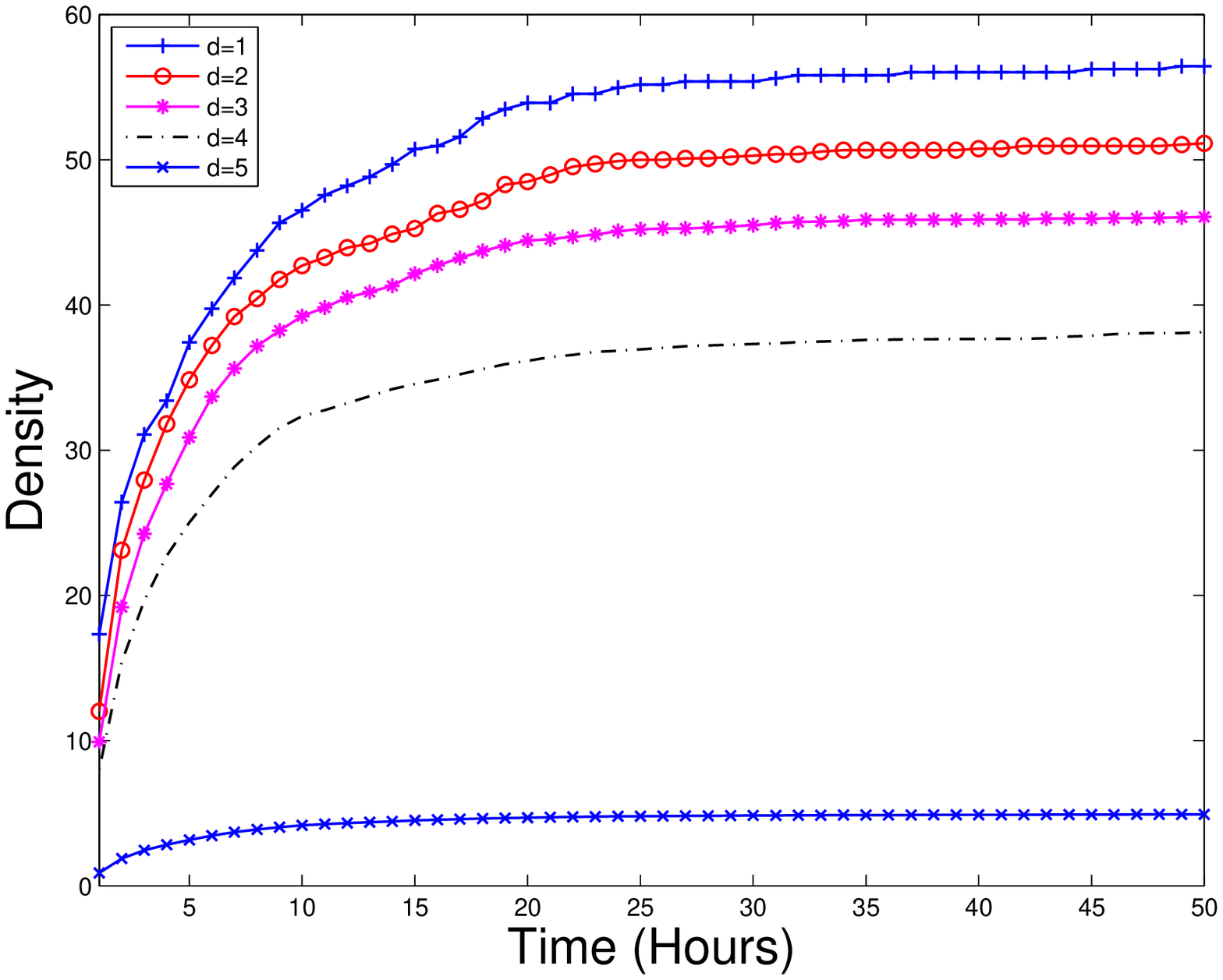}}
  \subfigure[Density of influenced users of s2]{
    \includegraphics[width=2.5in]{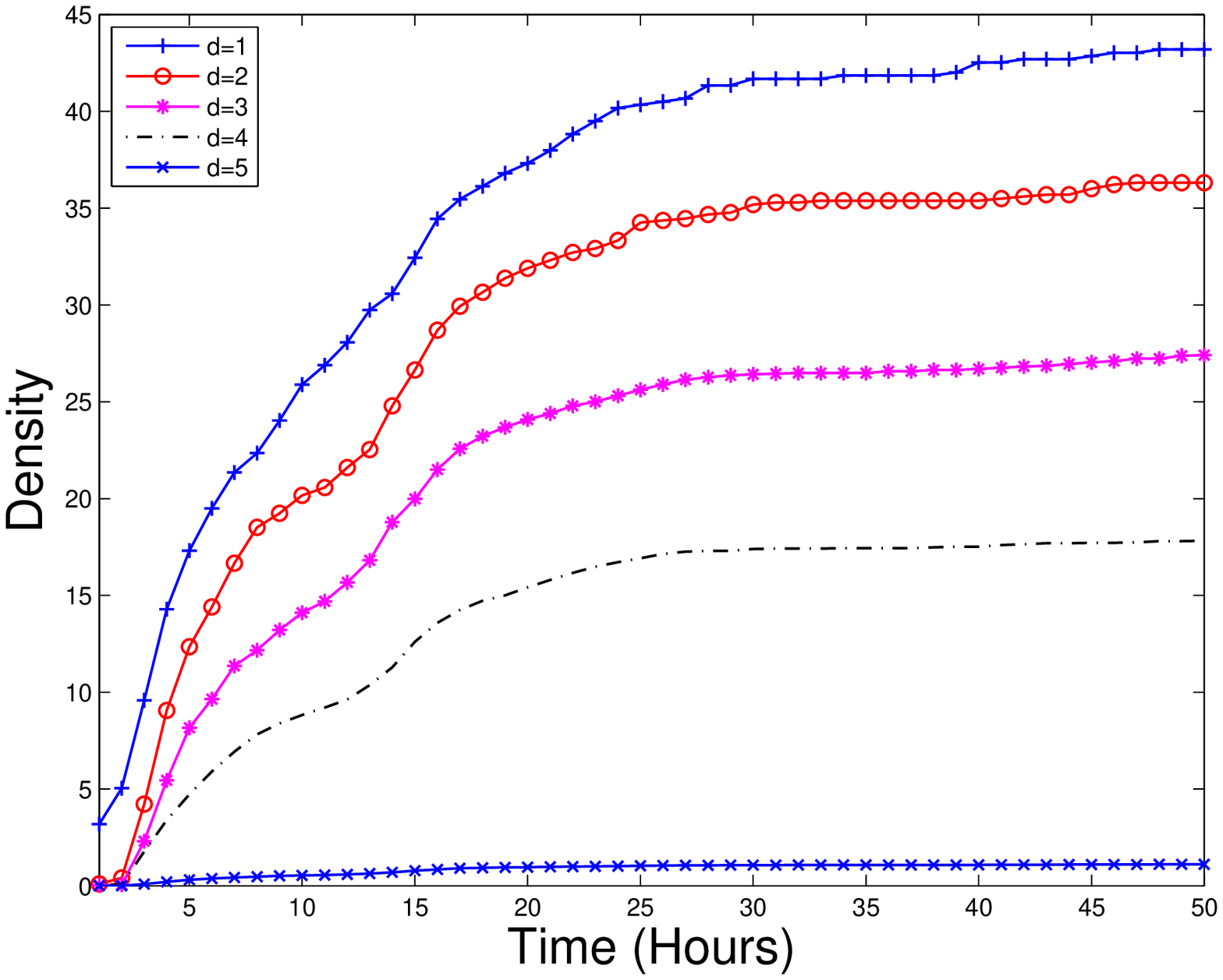}}
  \subfigure[Density of influenced users of s3]{
    \includegraphics[width=2.5in]{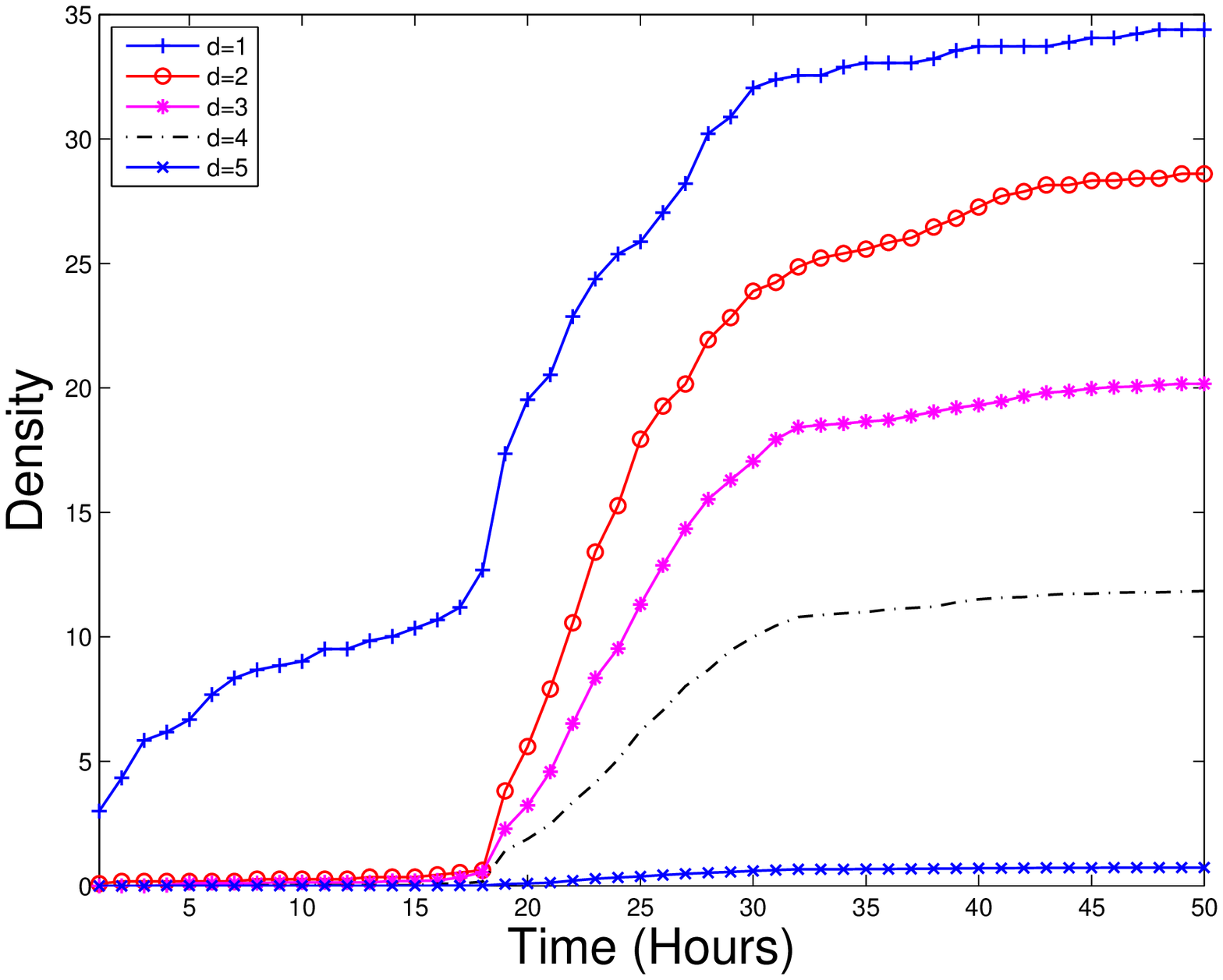}}
  \subfigure[Density of influenced users of s4]{
    \includegraphics[width=2.5in]{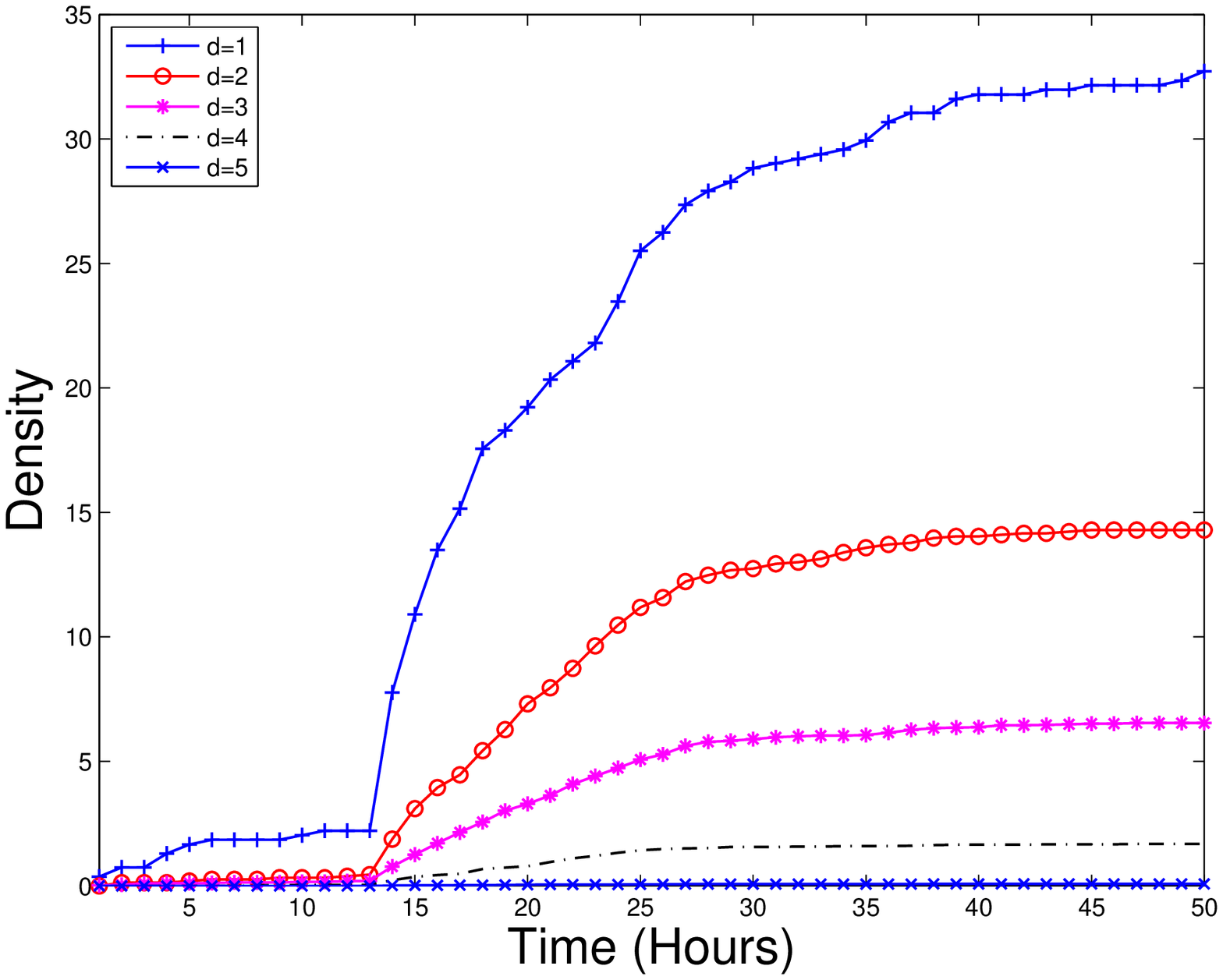}}
 \caption{Density of infected users over 50 hours with shared interests as distance}\label{fdensity_interest}
\end{figure*}

Next we study the impact of the distance based on shared interests on the density
of influenced users. For each top news story, we first calculate the shared interests distance
between the initiator and all other users, and classify the users into five disjoint
groups based on their interest ranges. To make the distance values consistent with friendship hops, we
assign value $1-5$ to each of the 5 groups. Figure~\ref{fdensity_interest} shows the density of influenced users
with shared interests as distance over the initial 50 hours for the same
four news stories as those in Figure~\ref{fdensity}, respectively. Interestingly, all four stories show the consistent pattern, i.e., as shared interests distance increases, the density of influenced users decreases. This confirms that interest plays an important role in information spreading in online social networks and the shared interests distance serves as a good distance metric.

\subsection{Predicting Information Diffusion with the DL model}
The empirical study shows that the density of influenced users of different stories exhibit different temporal and spatial patterns.
The difference could be caused by a variety of reasons, such as the influence of the initiator or the real interest
of the actual news story. Such difference makes it more challenging to predict information diffusion.  In this subsection, we evaluate the DL model given the initial spreading phase of a story.

The construction of initial density function follows the method
outlined in Section II.D. Specifically, we create the initial density function for each news story
using the density of influenced users captured at the first hour after the story was initially
posted on Digg. The carrying capacity $K$ is set to $25$ since in Figure~\ref{fdensity}[a], we observed that the density of influenced user of story s1 is always below $25$.
The diffusion rate $d$ is set to $0.01$. Based on the observations in Figure~\ref{f332346.hop.distance}, we set growth rate as a decreasing function of time $t$.
Growth rate $r(t)$ is defined in Equation~\ref{choicer} and illustrated in Figure 6.

\begin{equation}\label{choicer}
r(t) = 1.4e^{-1.5(t-1)} + 0.25
\end{equation}

\begin{figure}\label{choiceofr}
\begin{center}
  \includegraphics[width=6cm]{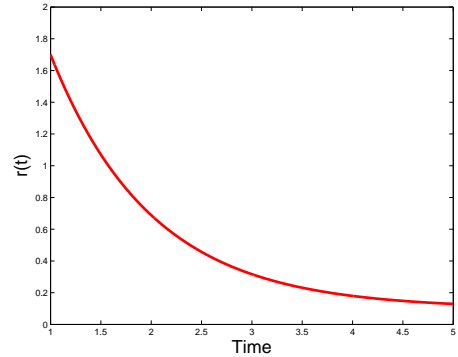}\\
  \caption{Growth rate $r$ is a decreasing function of time $t$}
 \end{center}
 \end{figure}

\begin{figure}[ht]
  \centering
  \subfigure[Predicted vs. Actual data of s1 with friendship hops as distance]{
    \includegraphics[width=2.5in]{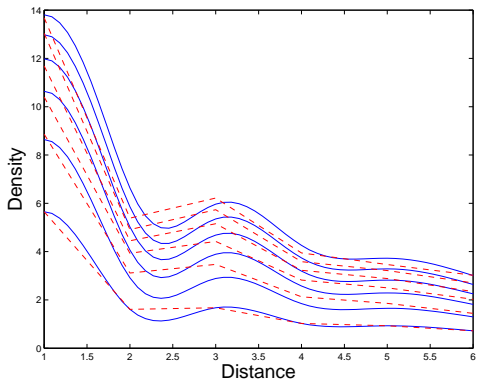}}
  \subfigure[Predicted vs. Actual data of s1 with shared interests as distance]{
    \includegraphics[width=2.5in]{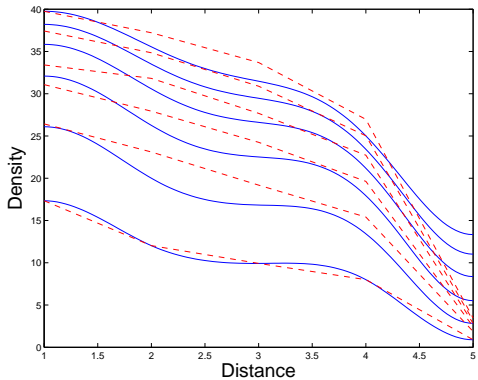}}
 \caption{Predicted vs. Actual data of story s1}
 \label{f332346.match}
\end{figure}

Figure~\ref{f332346.match}[a] illustrates the predicted results for the most popular story $s1$. $x$-axis is the distance measured by {\em friendship hops} and
$y$-axis represents the density of influenced users. The dashed lines denote the {\em actual} observations for the density at different times,
while the solid lines illustrate the density {\em predicted} by the DL model. Note that the lowest line representing $t=1$ is the initial density function. In online social networks, the density is only meaningful when distance is integer. It is clear that the predicted values closely match the actual values over time and distance. To quantify the accuracy of the prediction, we calculate the prediction accuracy (defined in Equation~\ref{accuracy}) of the DL model as follows:

\begin{equation}\label{accuracy}
\text{Prediction accuracy} = \frac{|\text{predicted value} - \text{actual value}|}{\text{actual value}}
\end{equation}

Table~\ref{tab:prediction} gives the average accuracy for users of distance $1$ to $6$ of the most popular story $s1$. It is calculated from the same data illustrated in Figure~\ref{f332346.match}[a]. As we can see, with the setting of initial density function, $r$, $K$ and $d$, the DL model gives most accurate prediction for users at distance 1 where the average prediction accuracy is $98.27\%$. The overall average prediction accuracy across all distances is $92.81\%$, which indicates the effectiveness of the DL model.
\begin{table}[htbp]
\centering
\scriptsize
\caption{The prediction accuracy with friendship hop as distances for story $s1$}
\label{tab:prediction}
\begin{tabular}{|c|c|c|c|c|c|c|} \hline

Distance & Average & t = 2 & t = 3 & t = 4 & t = 5 & t = 6 \\ \hline

1 & 98.27\% & 97.47\% & 97.74\% & 97.48\% & 99.55\% & 99.09\% \\ \hline

2 & 86.99\% & 93.59\% & 96.63\% & 87.16\% & 80.80\% & 76.78\%  \\ \hline

3 & 90.28\% & 83.23 \% & 87.98\% & 90.99\% & 93.35\% & 95.94\%  \\ \hline

4 & 92.98\% & 86.75\% & 91.39\% & 99.00\% & 95.68\% & 92.06\%  \\ \hline

5 & 93.77\% & 89.05\% & 91.61\% & 97.79\% & 97.92\% & 92.49\%  \\ \hline

6 & 94.56\% & 90.03\% & 89.48\% & 96.04\% & 97.57\% & 99.67\%  \\ \hline

\end{tabular}
\end{table}

We further validate the proposed model with the distance metric {\em
shared interests}. Similarly, the initial density function is constructed from the data collected at $t=1$. $d$ is $0.05$, $K$ is $60$, and $r(t)=1.6e^{-(t-1)}+0.1$.
The result is illustrated in Figure~\ref{f332346.match}[b]. The DL model gives close predictions for users at distance $1$ to $4$, except for users at distance $5$. As can be seen in Figure~\ref{f332346.match}[b], the actual density of influenced users at distance $5$ drops faster at time $2$ to time $5$ than at time $1$. This scenario tells us that the model can be refined by choosing a function of both distance and time for growth rate $r$, which we will explore as future work.

\begin{table}[htbp]
\centering
\scriptsize
\caption{The prediction accuracy with shared interests as distances for story $s1$}
\label{tab:predictioninterest}
\begin{tabular}{|c|c|c|c|c|c|c|} \hline
Distance & Average & t = 2 & t = 3 & t = 4 & t = 5 & t = 6 \\ \hline
1 & 97.21\% & 98.74\% & 96.75\% & 92.70\% & 97.91\% & 99.97\%  \\ \hline
2 & 93.67\% & 86.58\% & 93.99\% & 96.11\% & 96.14\% & 95.52\%  \\ \hline
3 & 93.11\% & 87.71\% & 92.86\% & 96.14\% & 95.39\% & 93.44\%  \\ \hline
4 & 91.64\% & 87.18\% & 91.38\% & 93.23\% & 93.63\% & 92.75\%  \\ \hline
5 & 39.84\% & 66.26\% & 44.43\% & 33.91\% & 28.68\% & 25.92\%  \\ \hline

\end{tabular}
\end{table}

In summary, the experiment results based on Digg datasets show that our
PDE-based DL model effectively characterizes and predicts the process of
information cascading over online social networks. Furthermore, the proposed
model works very well with distance metrics measured by either {\em friendship hops}
or {\em shared interests}.

\section{Related Work}
Information diffusion over online social networks has drawn much attention from the research community. Most prior work have focused on empirical studies in different online social networks. For example, ~\cite{Gruhl:2004} studied the dynamics of information propagation in weblogs. ~\cite{ChaWOSN08,Cha09, YuFSKD09} studied cascade patterns
of disseminating popular photos over Flickr social network, while~\cite{LermanICWSM10}
examined how the interest in news stories spreads among the users in
Digg and Twitter social networks based on empirical data
extracted from these networks. ~\cite{SteegPS11} studied the factors that prohibit the epidemic transmission of popular news posted on Digg. In addition, Tang {\em et al.} presented a large-scale
empirical study on network structure, user characteristics, and content dissemination process
of Digg social network~\cite{TangTM11}.

Several recent work have proposed mathematical models for understanding and predicting
information diffusion in online social networks over time. \cite{Kazumi11} characterized
the process of information diffusion over social networks during a given
time period using SIS (Susceptible, Infected, and Susceptible) epidemic model.
An earlier work~\cite{KempeKDD03} used two most basic diffusion models, namely
{\em Linear Threshold} and {\em Independent Cascade Models}, in searching the most
influential users in online social networks. In~\cite{YangICDM10}, Yang {\em et al.}
introduced a {\em Linear Influence Model} to predict the number of newly infected nodes based on the time when previous set of nodes are infected.
~\cite{GomezRodriguez10} proposed a model to infer underlying paths of information diffusion, while~\cite{Budak11} studied how to limit the spread of misinformation.

In addition, several research work have been done to study the impact of
the structure of online social networks on the process of information diffusion.
In particularly, \cite{YangWSM10} studied the information structure of Weblogs
and microblogs and revealed their systematical difference in contribution
and navigation patterns and user interactions, while \cite{Harvard09}
examined the impact of human activity patterns on the dynamics
of information diffusion using a viral email experiment involving
over 31 thousand individuals. Further, \cite{KatonaJMR11} demonstrated the effects of users' connection patterns in an online social
network on the information diffusion process.

Our work is different from these prior studies in online social networks because we focus on mathematical models to predict the information diffusion in both temporal and spatial dimensions in online social networks, which has not been addressed in any of these studies.

There are also parallel research in biology, sociology, economics, and physics to model time evolution systems~\cite{Cantrell2003, Fife1979}. Many of them use PDE model which provides a way to translate local assumptions or data about the movement and reproduction of individuals into global conclusions about the population. However, the concepts we introduced in this paper such as distance, diffusion, and growth are abstract and unique for online social networks.

\section{Conclusions and Future Work}
This paper introduces a novel PDE-based diffusive logistic model for solving
the {\em spatio-temporal diffusion problem} in online social networks. Through
measuring the density of influenced users at a certain distance from the source
of the information during a given time period, we characterize the temporal and
spatial patterns of information diffusion process. Further, we use the proposed DL model to predict the density of influence
users with varying distances over time based on the early phase of information
spreading activities, and evaluate the prediction quality using datasets
collected from Digg social networks. Our experiments results show that the model
achieves a prediction accuracy of over 92\% on the density of influenced users for
the representative news stories during the first six hours since these stories were
submitted. Our future work lies in developing new models that consider diffusion
rate, growth rate and carrying capacity as functions of time and distance. In addition,
we plan to investigate the effectiveness of the DL model on predicting
the spatio-temporal diffusion process over other social networks such
as Facebook and Twitter.

\end{document}